\documentclass{sig-alternate-2013}
\usepackage{ctable}
\usepackage{graphicx}
\usepackage{algorithmic}
\usepackage[hyphens]{url}
\usepackage{tabularx}
\usepackage{times}
\usepackage[utf8]{inputenx}
\usepackage{textcomp}
\usepackage{balance}
\usepackage{subcaption}
\usepackage{tabu}
\usepackage{dblfloatfix}
\usepackage{enumitem}
\usepackage{verbatim}
\usepackage{morefloats}

\usepackage{wrapfig}

\permission{Copyright is held by the International World Wide Web Conference Committee (IW3C2). IW3C2 reserves the right to provide a hyperlink to the author's site if the Material is used in electronic media.}
\conferenceinfo{WWW'14 Companion,}{April 7--11, 2014, Seoul, Korea.} 
\copyrightetc{ACM \the\acmcopyr}
\crdata{978-1-4503-2745-9/14/04. \\
http://dx.doi.org/10.1145/2567948.2576943
}

\begin{document}
%



\title{Evolution of Reddit: From the Front Page of the Internet to a Self-referential Community?}
\makeatletter
\renewcommand*{\@fnsymbol}[1]{*}
\makeatother

\numberofauthors{5}
\author{
   \alignauthor Philipp Singer\thanks{Both authors contributed equally to this work.} \\
     \affaddr{\mbox{Graz University of Technology}}\\
     \email{\mbox{philipp.singer@tugraz.at}}\\
     \and
   \alignauthor Fabian Flöck\footnotemark[1]\\
     \affaddr{\mbox{Karlsruhe Institute of Technology}}\\
     \email{\mbox{floeck@kit.edu}}\\
   \and
   \alignauthor Clemens Meinhart\\
     \affaddr{\mbox{Graz University of Technology}}\\
     \email{\mbox{c.meinhart@student.tugraz.at}}\\
     \and
     \alignauthor Elias Zeitfogel\\
     \affaddr{Graz University of Technology}\\
     \email{elias.zeitfogel@student.tugraz.at}\\
     \and
     \and
   \alignauthor Markus Strohmaier\\
     \affaddr{\mbox{GESIS \& U. of Koblenz}}\\
     \email{\mbox{markus.strohmaier@gesis.org}}\\
}
\maketitle


\maketitle

\raggedbottom
\interfootnotelinepenalty=10000

\begin{abstract}



In the past few years, Reddit -- a community-driven platform for submitting, commenting and rating links and text posts --
has grown exponentially, from a small community of users into one of the largest online communities on the Web. 
To the best of our knowledge, this work represents the most comprehensive longitudinal study of Reddit's evolution to date, studying both (i) how user submissions have evolved over time and (ii) how the community's allocation of attention and its perception of submissions have changed over 5 years based on an analysis of almost 60 million submissions. Our work reveals an ever-increasing diversification of topics accompanied by a simultaneous concentration towards a few selected domains both in terms of posted submissions as well as perception and attention. By and large, our investigations suggest that Reddit has transformed itself from a dedicated gateway to the Web to an increasingly self-referential community that focuses on and reinforces its own user-generated image- and textual content over external sources. 

\end{abstract}

\category{H.3.5}{Information Storage and Retrieval}{Online Information Services}


\keywords{Reddit; frontpage; evolution; growth; longitudinal; online community; attention; perception; discussion; diversification; self-reference}

\section{Introduction}
\label{sec:intro}

\begin{figure}[t!]
\vspace{-0.2cm}
\centering
\includegraphics[width=0.85\columnwidth]{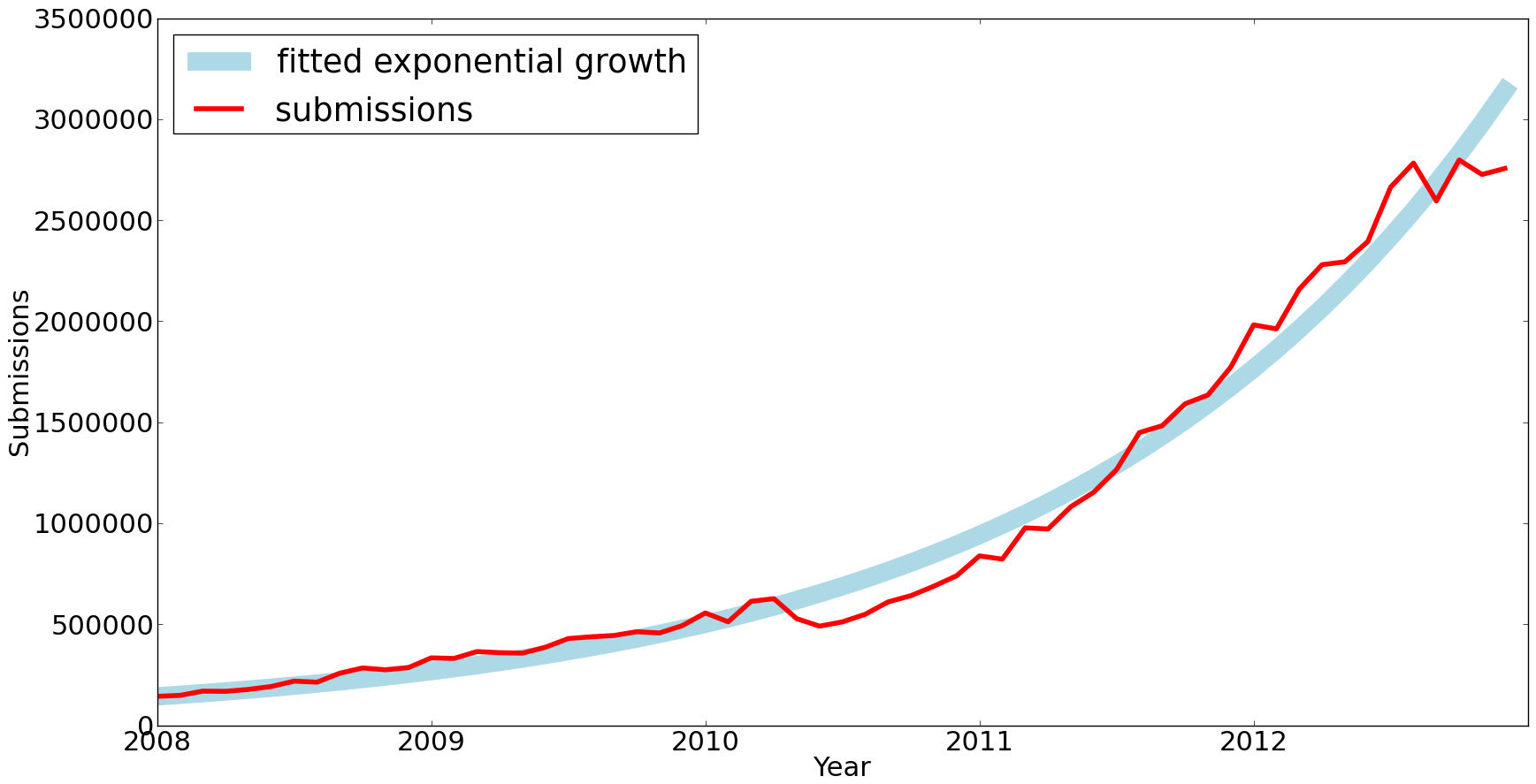}
\caption{Number of submissions to Reddit each month, ranging from Jan. 2008 to Dec. 2012 (red line). The blue line denotes the best exponential fit for the growth, ongoing until the end of 2012.}
\label{fig:exponential}
\vspace{-0.2cm}
\end{figure}

Since its founding in 2005, Reddit 
has grown into one of the largest online communities on the web. As of this writing, the site has more than 112 million unique visitors from over 195 countries each month.\footnote{\label{reddit/about}\url{http://www.reddit.com/about/}, as of Feb. 02$^{th}$, 2014} It is ranked by \emph{Alexa.com} as the 69$^{th}$ and 27$^{st}$ most popular website in the world and the U.S., respectively.\footnote{\url{http://www.alexa.com/siteinfo/reddit.com}, as of Feb. 02$^{th}$, 2014}
On Reddit, users can post links to external websites \emph{or} submit textual content directly hosted on Reddit, so-called self submissions or self posts.
Other ``Redditors'' -- a neologism combining ``Reddit'' and ``editor'' -- can then up- and downvote the posted items, contributing to an ever-changing ranking of the ``hottest'' submissions.
Users can comment on every submission as well as create their own sub-communities named ``subreddits'' -- each being independent, dedicated to a specific topic and moderated by volunteers.
Equipped with these features, Reddit was intended to capture and rank all kinds of diverse content collected from the Web by promoting the best parts via its voting process. Reddit's original claim is that the site represents \emph{"the front page of the Internet"} -- suggesting that it acts as a gateway to (the best) content available on the Web. Today, this declaration is still prominently featured in the HTML title of \emph{Reddit.com}. 
With this mission statement, the platform has exhibited exponential growth in terms of submissions between 2008 and 2012, as evident in Figure~\ref{fig:exponential}.\footnote{Exponential growth being a better fit than a Gompertz model as well as a logistic  model, tested on our data described in Section \ref{sec:dsmethods}.} 
Yet, 
it remains for the most part unclear whether the initial design intentions behind Reddit are still relevant today, or whether the system has evolved to accommodate other purposes. In the following we aim to address this question.




\textbf{Research questions:} 
Specifically, we address two issues:\\
\emph{(i) Longitudinal analysis of user submissions:} We examine in detail how user submissions to Reddit have evolved over the course of five years. Regarding diversity of subreddits, top-level domains of posted links and types of content allow us to evaluate whether and how the focus of user posts to Reddit has changed over time. 
\emph{(ii) Longitudinal analysis of perception and attention:} To gauge whether and how perception and attention by the Reddit community developed, we analyze voting and commenting patterns, enabling us to assess what kind of submissions received attention by Redditors over time. 
In this work, we use a large-scale dataset containing all submissions to Reddit 2008-2012 (close to 60 mio submissions). A succinct user survey supplements our analysis.



\textbf{Contributions \& results}: To the best of our knowledge, this work represents the most comprehensive longitudinal study of Reddit's evolution to date, studying both (i) how user submissions have evolved over time and (ii) how the community's allocation of attention and its perception of submissions have changed over 5 years based on an analysis of almost 60 million submissions. Our analysis of all Reddit submissions from 2008-2012 reveals an ever-increasing diversification of topics (i.e., subreddits) accompanied by a simultaneous concentration towards a few selected domains and types of submissions (self and image). This suggests that Reddit has transformed itself from a dedicated gateway to Web content to an increasingly diverse, self-referential community that focuses on and reinforces its own user-generated content over external sources.
Our work sheds light on formerly unknown dynamics of Reddit and represents an important step towards a deeper understanding of Reddit and similar platforms. 


\textbf{Structure of this paper:} In the next section, we describe our dataset and methods.  We present our results in Sections~\ref{sec:diversity} and \ref{sec:perception} and discuss them in Section~\ref{sec:discussion}. After listing related work in Section~\ref{sec:related}, conclusions and future work are summarized in Section~\ref{sec:conclusions}.

\section{Dataset and Methods}
\label{sec:dsmethods}
In this section, we introduce our dataset\footnote{Dataset access can be requested at \url{http://www.philippsinger.info/reddit/}.} and describe the method used for categorizing the content submitted to Reddit into six categories, before explaining the design of our user survey. 

\subsection{Description of the dataset}
\label{subsec:dataset}

We analyze data consisting of all submissions posted to Reddit from January 2008 to December 2012 crawled through Reddit's API.\footnote{\url{http://www.reddit.com/dev/api}} The metadata of each submission (i.e., title, author, up- and downvotes, number of comments, the link or text it contained and the submission time) were collected
around 1-2 months after the initial submission (i.e., when they get blocked from voting) as the metadata has most likely been settled after this period. 
Overall, we analyze 58,874,22 submissions (14,979,707 self posts) in 125,662 distinct subreddits from 4,910,850 different authors linking to 1,841,239 distinct domains on the Web.

\textbf{Categorizing submissions on Reddit:} 
We also provide a categorization of the content of the links that are submitted for facilitating an analysis of the types of content on Reddit.
We manually classified the 100 most frequently submitted domains which represent 69\% of all submissions (including self posts), into six categories: \textit{text, image, video, audio} and \textit{misc} and the last category \textit{self}, which accounts for all self posts in the dataset (25\% on their own). 
Domains were assigned a single category after examining their main purpose or type of content. The \textit{text} category covers everything from news-sites to blogs with focus on textual content and even encyclopedias (e.g., Wikipedia). \textit{Image, video} and \textit{audio} are mainly hosting services and content providers for specific types of media; the most used examples in the dataset would be \emph{Imgur.com, Youtube.com and Soundcloud.com}, respectively. The \textit{misc} category covers domains that do not clearly fit into one of the other categories and comprises, e.g., link shorteners like \emph{Tinyurl.com} or universal hosting services like \emph{Amazon Web Services}. 


\subsection{Description of the user survey}
\label{sec:usersurvey}

After the main data collection, a short auxiliary user survey with questions regarding certain aspects of this paper was posted to the subreddits \emph{r/theoryofreddit}  and \emph{r/samplesize}, as their user communities are very open to providing answers to questionnaires.\footnote{\label{fn:survey}The full questions plus additional information can be found at\\ \url{http://people.aifb.kit.edu/ffl/redditsurvey}} This particular, limited sampling and the self-selection of respondents must be taken into account when interpreting the results. Our analysis showed, however, no notable difference between the answer patterns of the two subreddits and will henceforth be reported in aggregate. 
The survey ran from Nov. 24 until Dec. 1, 2013 and yielded 1,004 responses (Note: some questions were optional and not answered by all users). We filtered obvious spam answers from the results, leaving 969 answers: 66\% from \emph{r/theoryofreddit} and 34\% from \emph{r/samplesize}.\footnote{Unreasonable values (like ``25 hours per day'') and text (nonsensical, flaming) were used as indicators of spam.} 
Questions and results from the survey will be reported as supplemental information in selected sections.



\section{Diversity and self reference}
\label{sec:diversity}

As Reddit has experienced exponential growth (cf. Section~\ref{sec:intro}), it is not unlikely that the internal dynamics of Reddit have evolved correspondingly and thereby affected the character of the site as a whole and particularly its function as the ``front page of the Internet''. 
We are thus interested in investigating the current distribution and change over time of three important aspects of Reddit: (a) subreddits, (b) linked domains and (c) types of linked content.

\subsection{The diversification of subreddits}
\label{sec:changesubreddit}


\begin{figure*}[h!t!]
\vspace{-0.3cm}
\centering
\begin{subfigure}[b]{0.33\textwidth}
\includegraphics[width=\textwidth]{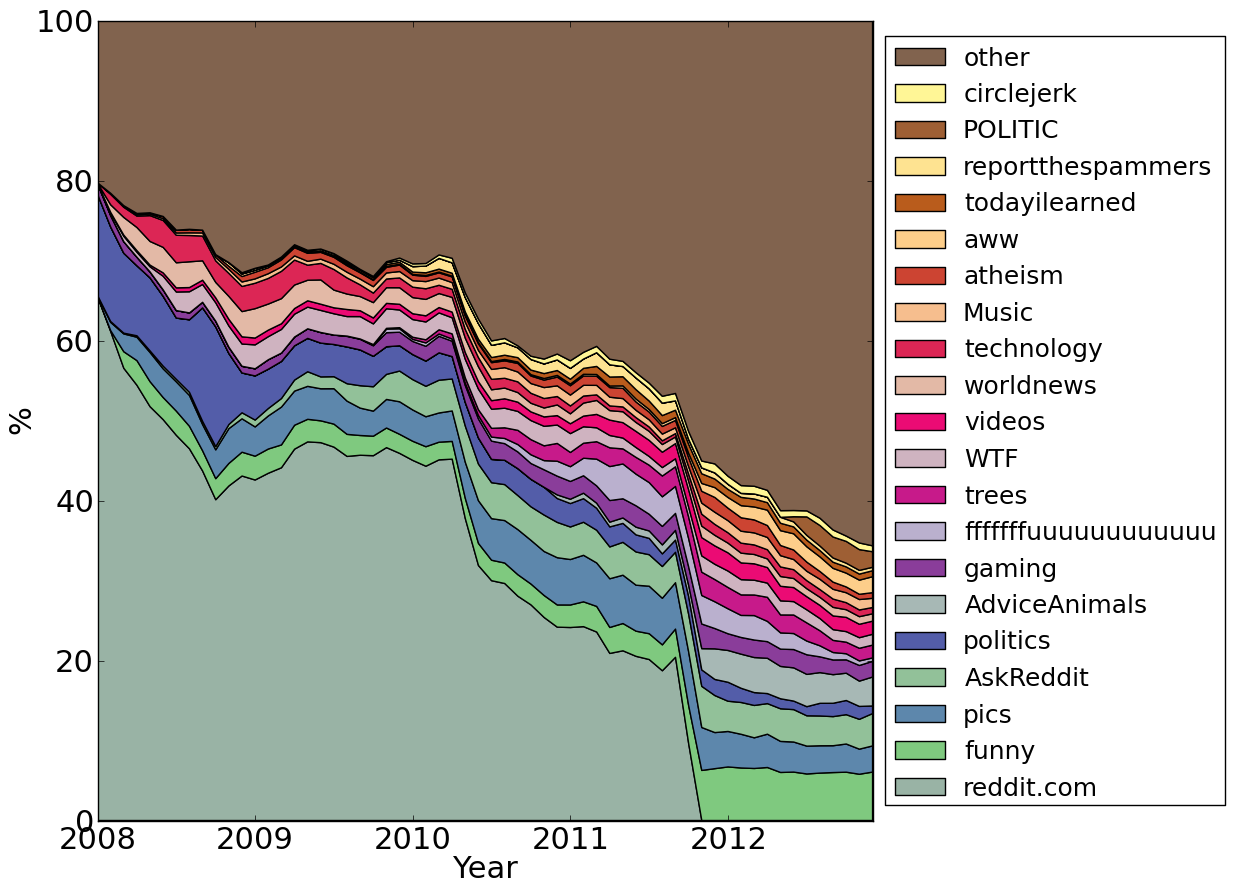}
\caption{Subreddits}
\label{subfig:subredditchange}
\end{subfigure}
\begin{subfigure}[b]{0.33\textwidth}
\includegraphics[width=\textwidth]{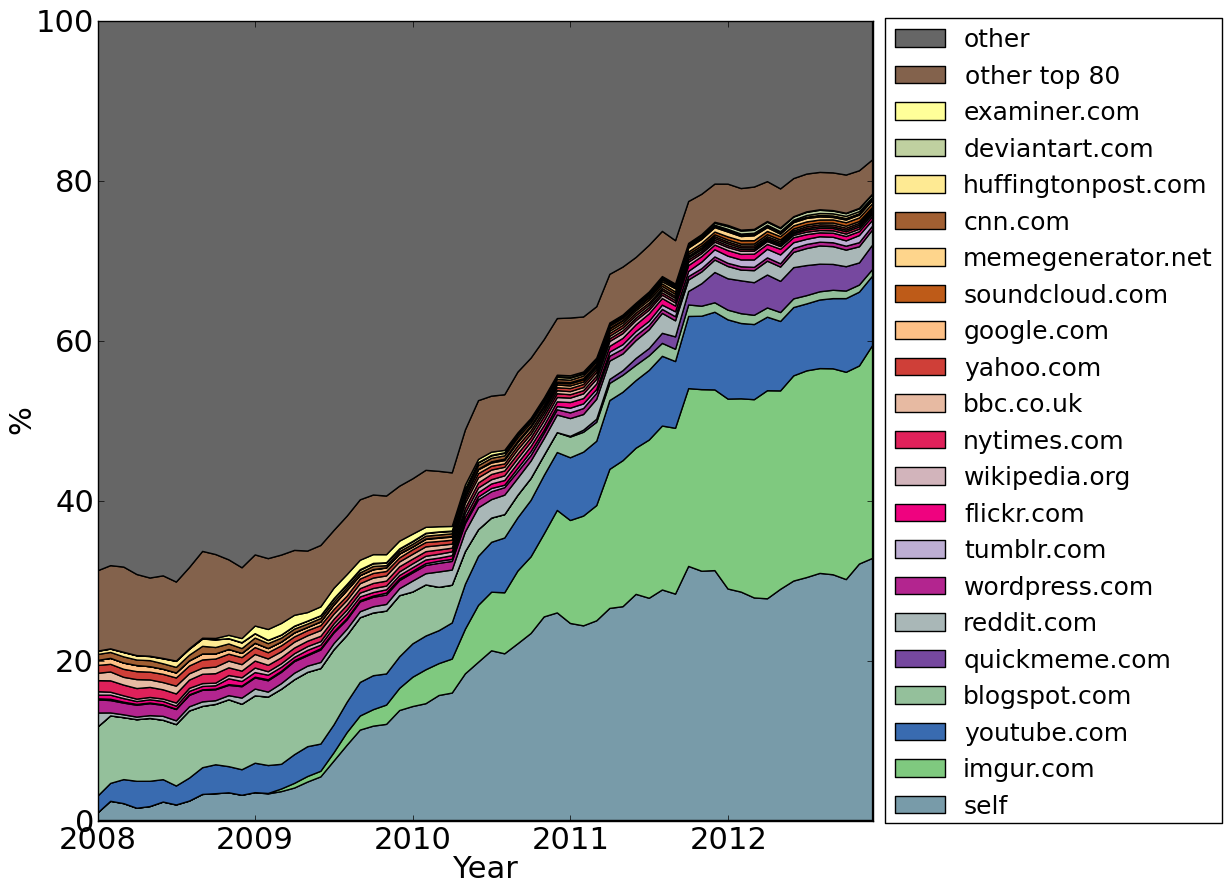}
\caption{Domains}
\label{subfig:domainchange}
\end{subfigure}
\begin{subfigure}[b]{0.33\textwidth}
\includegraphics[width=\textwidth]{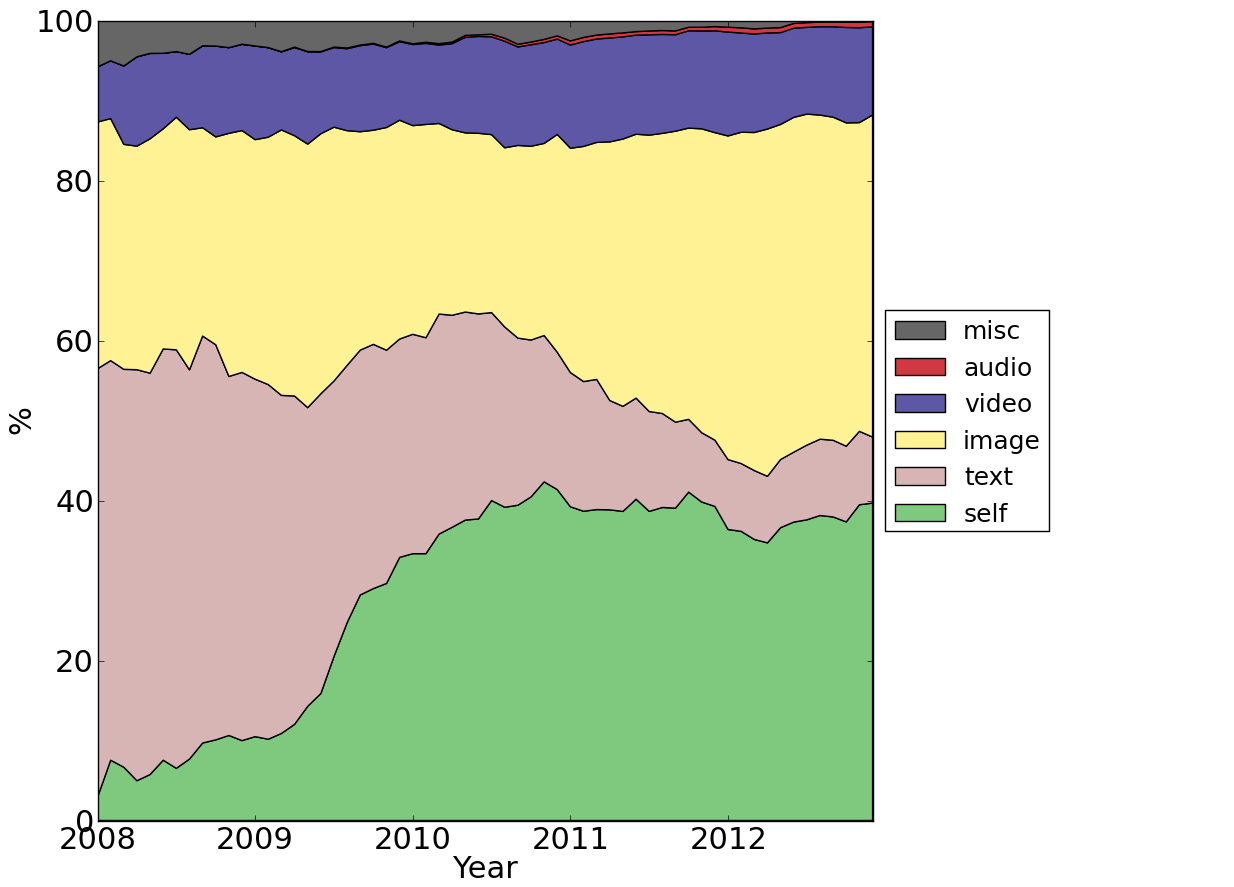}
\caption{Type of content}
\label{subfig:contentchange}
\end{subfigure}
\caption{Evolution of submissions per month over subreddits, domains and types of content, 2008-2012.}
\label{fig:subredditdomaincontentgrowth}
\end{figure*}

As outlined in Section~\ref{sec:intro}, one main aspect of Reddit today is the existence of thousands of distinct -- mostly user-created -- sub-communities, also called subreddits. 
We measure the popularity of subreddits by counting committed submissions to a particular subreddit at a specific time (monthly). 
In Figure~\ref{subfig:subredditchange} the development of all active subreddits (i.e., with at least one submission) is depicted, with their relative size in percent compared to the overall size in total submissions on Reddit at a specific time. We only visualize the 20 largest subreddits with distinct colors and combine the rest in brown color. 
We can observe that a fragmentation of submissions into an ever-increasing amount of distinct subreddits has taken place since Reddit's inception. 
In the last month of our data set in 2012, 32,202 subreddits received one or more submissions while  only 213 did so at the beginning of 2008. The 20 biggest subreddits at the end of 2012 contained less than 40\% of all submissions to Reddit, while they contained around 70\% and 80\% in mid-2010 and mid-2008, respectively. Measured as a Gini coefficient, the concentration of submissions over subreddits decreased slightly from 0.97 in mid-2008, over 0.95 in mid-2010 to 0.94 at the end of 2012.
These findings point, at first glance, to a strong diversification of topics represented by the different subreddits, whose establishment does in fact articulate the more explicit need of a part of the user base for certain dedicated thematic spaces. However, many topics and discourses might have existed previously as part of one of the broader themed subreddits, especially \emph{r/Reddit.com}, which served as the default posting space in the early phase of Reddit.
Figure~\ref{subfig:subredditchange} unveils  \emph{r/Reddit.com's} gradual demise. When user-founded subreddits were first introduced at the beginning of 2008, a great quantity of submissions was still committed to \emph{r/Reddit.com}.$^{\ref{reddit/about}}$ With more subreddits founded, \emph{r/Reddit.com} kept shrinking, until in October 2011 a set of 20 default subreddits was introduced, which led to \emph{r/Reddit.com's} deliberate shutdown by the operators.\footnote{\url{http://blog.reddit.com/2011/09/independence.html}} 

Overall, an at least \emph{structural diversification} can be noted in the form of a vast increase in subreddits; alas, it cannot be stated with certainty that the general \emph{thematic diversity of submissions} has in fact increased.
Some topics might have just been outsourced from more general subreddits similar themed subreddits to sharpen their profile and/or not to get lost in the increasing flood of posts in \emph{r/Reddit.com}. 
What can be affirmed, however, is that clearly distinct communities around topics had a chance to form in the secluded spaces of the subreddits, each with their own, clear-cut rules and ``submission ethics'' -- visible at the right-hand sidebar of most subreddits. 
In sum, the subreddit diversification fits sufficiently well with the claim of Reddit representing (the best) content from all over the Web, as the ladder grew exponentially in its already vast content heterogeneity since 2008 and Reddit has seemingly be able to mirror this diversity and built sub-communities around it.

\subsection{Towards self-reference in submissions and their linked domains}
\label{sec:changedomains}

On Reddit users can submit content (a) through self (text) submissions that are stored on Reddit itself and (b) through links to external Web content. The majority of submissions has traditionally been links to external content, as the initial idea of the platform was solely to share links and vote on them. 
For Reddit to be in fact a ``frontpage'' for -- or a gateway to -- content from all over the Web, one would expect an increasing diversity among the link targets (i.e., top level domains) of the submissions similar to what we have seen in the subreddit distribution. 
Figure~\ref{subfig:domainchange} reveals that the relative proportion of self submissions has been growing tremendously over time, with an initial boost in mid-2009. A small increase in backlinks to reference older submissions on \emph{Reddit.com} is also visible until mid-2010 and stabilizes at around 2\% afterwards. 
Apparently, the biggest external beneficiary of the content expansion is one image hoster called \emph{Imgur.com} (0\% in mid-2008, 7.32\% in mid-2010 and 26.6\% at the end of 2012) which owes its uprising to the large influx of Reddit posts.\footnote{\url{http://www.mediaite.com/online/imgur-accounts-alan-schaaf-interview/}} 
Imgur was in fact created for the expressed purpose of serving as an image hosting service for Reddit; Imgur founder Alan Schaaf stated in 2009 that he designed the platform as 
he was not satisfied with other available image hosts.\footnote{\url{http://www.reddit.com/tb/7zlyd/}}
Since its inception, Imgur is has not only risen to be the primary image host for Reddit, it has actually become a central aspect of Reddit's culture; so much in fact, that it is close to being the sole image hosting option for Reddit posts that is accepted by many subreddits, as community discussions reveal.\footnote{\label{imgurdiscuss}Cf. discussion threads on \url{http://www.reddit.com/r/TheoryOfReddit/search?q=imgur&restrict_sr=on}} 

The external domain exhibiting the second largest expansion in linked content is \emph{Youtube.com} (2.37\% in mid-2008, 6.24\% in mid-2010 and 8.68\% at the end of 2012), followed by \emph{Quickmeme.com} (0\% in mid-2008, 0\% in mid-2010 and 3.05\% at the end of 2012), a website providing a an easy-to-use service for creating meme images with custom personal messages. 
The two domains that suffered most were \emph{Blogspot.com} (7.68\% in mid-2008, 3.03\% in mid-2010 and 0.83\% at the end of 2012) and \emph{Wordpress.com} (1.41\% in mid-2008, 0.99\% in mid-2010 and 0.49\% at the end of 2012), both hosting mainly blogs, i.e., text based content.

Thus, while Section~\ref{sec:changesubreddit} unveiled that the content on Reddit frays out into more and more subreddits -- i.e., thematic subspaces -- over time, we now see that submissions to external content concentrated more and more to just a few domains, mainly self and \emph{Imgur.com}. The Gini coefficient for concentration accordingly increased notably from 0.78 in mid-2008, over 0.83 in mid-2010 to 0.95 at the end of 2012 -- computed over URLs that received submissions that month. The overall diversity of linked domains has meanwhile been keeping up fairly in accordance with the submission growth, with the total number of distinct domains being 34,082 in mid-2008, 68,577 in mid-2010 and 103,660 at the end of 2012. The shifting focus on self-referential posts thus evolved parallel to an otherwise still diverse spectrum of linked domains. 


\subsection{A shift to ``self'' and images}
\label{sec:contentchange}

To better understand the distribution of links over external domains, we analyze the type of content these URLs usually host and how the popularity of each content type has progressed. To this end, we make use of the classifications of top-level domains provided in Section~\ref{subsec:dataset} into image, video, text, audio, misc and self. 

The progression over time -- represented via the relative proportion of each category, cf. Figure~\ref{subfig:contentchange} -- confirms  that self posts have not always been the favorite kind of submission.
From 2008 to mid-2009 the majority of submissions were linking to external textual content. Over time, the (likewise textual) self submissions exceeded the number of external textual submissions, consistent with the decline of \emph{Blogspot.com} and \emph{Wordpress.com}. 
Yet, while some material from blog sites or even news portals formerly linked might have ``migrated'' directly into self posts by the end of 2012, it is unlikely that all vanished text links are mirrored in self submissions.

Congruent
with the observations made in regard to \emph{Imgur.com} in Section~\ref{sec:changedomains}, image submissions have been growing. The increase in image posts, however, lags behind Imgur's expansion, which suggests that Imgur not only serves 
the supplemental need for picture storage Reddit has experienced since mid-2010 but in addition took over traffic from other image hosting sites.
$^{\ref{imgurdiscuss}}$
Next, we want to shed more light on the main topics of self and image submissions.





\begin{figure}[t!]
\centering
\includegraphics[width=0.86\columnwidth]{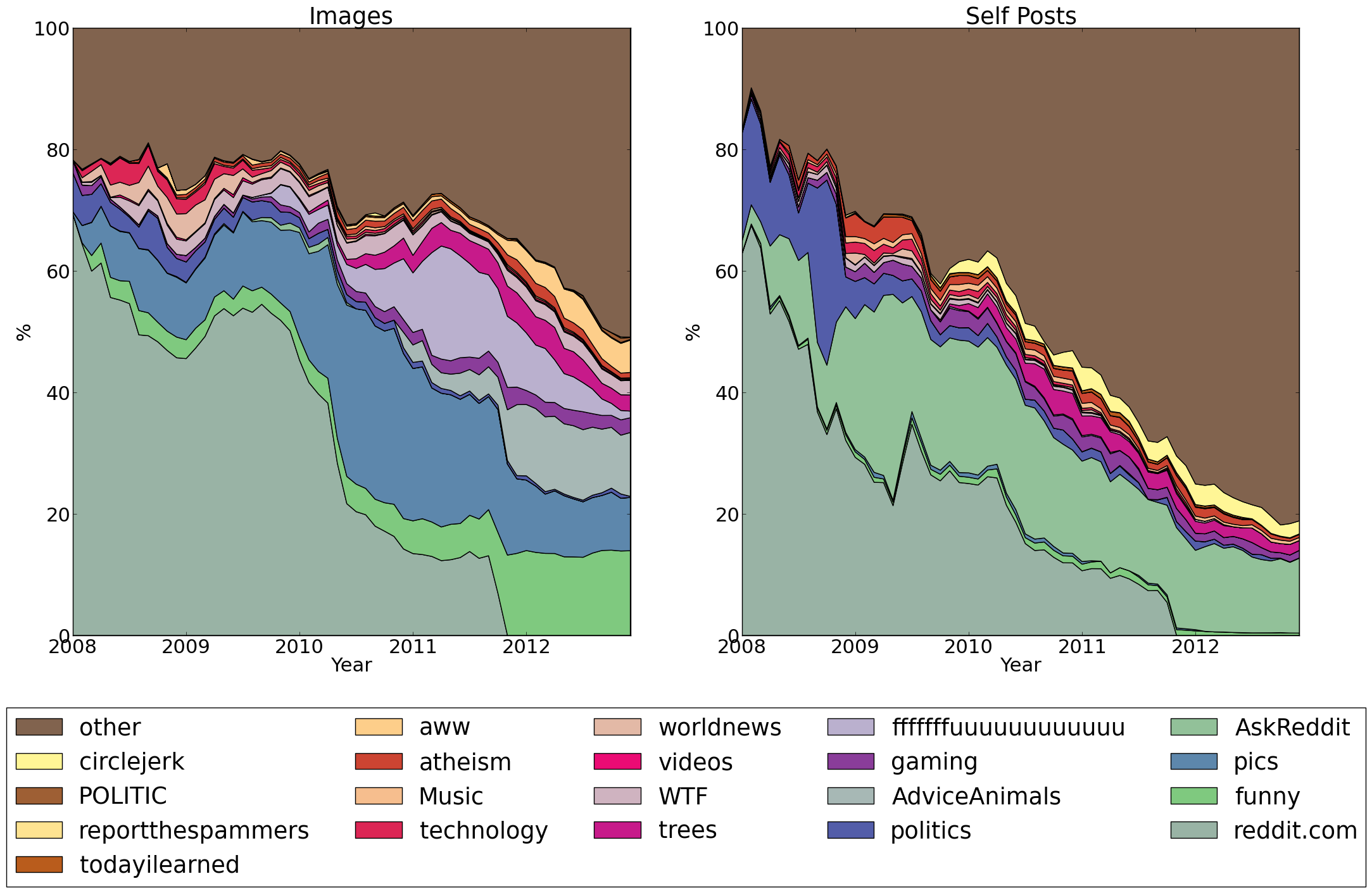}
\caption{Evolution of submissions per month over subreddits for image posts (left) and self posts (right), 2008-2012.}
\label{fig:imageselfoversubs}
\vspace{-0.4cm}
\end{figure}

At the end of 2012, we can identify an almost completely balanced ratio between textual content (self and text) and media content (image, video, audio). 
A closer look at the subreddits where image submissions are mainly posted (Figure \ref{fig:imageselfoversubs}) reveals that images have traditionally most frequently been used in \emph{r/Reddit.com} and, not surprisingly, \emph{r/pics}, a prominent subreddit dedicated to sharing any kind of interesting pictures (excluding adult content and images with superimposed text). Users post images they found around the Web in \emph{r/pics}, but to a very large extent submit photos they took themselves \emph{or} that tell a personal story.
When \emph{r/Reddit.com} died at the end of 2011, two subreddits saw a huge surge in image submissions. Firstly, \emph{r/funny}, a subreddit for, plainly, all things funny; it also features a large share of personal images, often superimposed with meme-like text. Secondly, \emph{r/AdviceAnimals}, dedicated to user-created memes of animal pictures which are combined with short, serious or (mostly) joking advice messages. 
Other subreddits remained relatively stable over time in regard to image posts, with the exception of \emph{r/fu} (abbreviation, original: \emph{r/fffffffuuuuuuuuuuuu})
which exclusively hosts a meme called ``rage comics'', packaging everyday stories into user-created comic strips of a certain format. This subreddit saw a profound upswing in image posts at the beginning of mid-2010, but leveled off towards the end of 2012.
By and large, it is a justified assumption that most image submissions from \emph{r/Reddit.com} moved to \emph{r/AdviceAnimals} and \emph{r/funny} upon its dismissal, while these subreddits kept growing further subsequently. 

As shown in Figure~\ref{fig:imageselfoversubs}, self posts have had their home as well mainly in \emph{r/Reddit.com} and -- at least 
until its slow fade-out starting in 2009 -- in \emph{r/politics}, the main subreddit for political matters.
While \emph{r/Reddit.com} has been in steady decline since 2008 (also) regarding self submissions, \emph{r/AskReddit} has, since mid-2008, quickly become the first and foremost place for self posts.  In this subreddit, users can post any ``open-ended, discussion-inspiring questions'' according to its self-description, that are in succession answered and discussed by the collective of Redditors. 
Another small but notable and stable increase in self submissions can be confirmed for \emph{r/circlejerk} since mid-2009, a satirical subreddit poking fun at typical habits of Reddit users and the community.

\section{Attention and perception}
\label{sec:perception}

\begin{figure*}[t!]
\vspace{-0.2cm}
\centering
\begin{subfigure}[b]{0.33\textwidth}
\includegraphics[width=\textwidth]{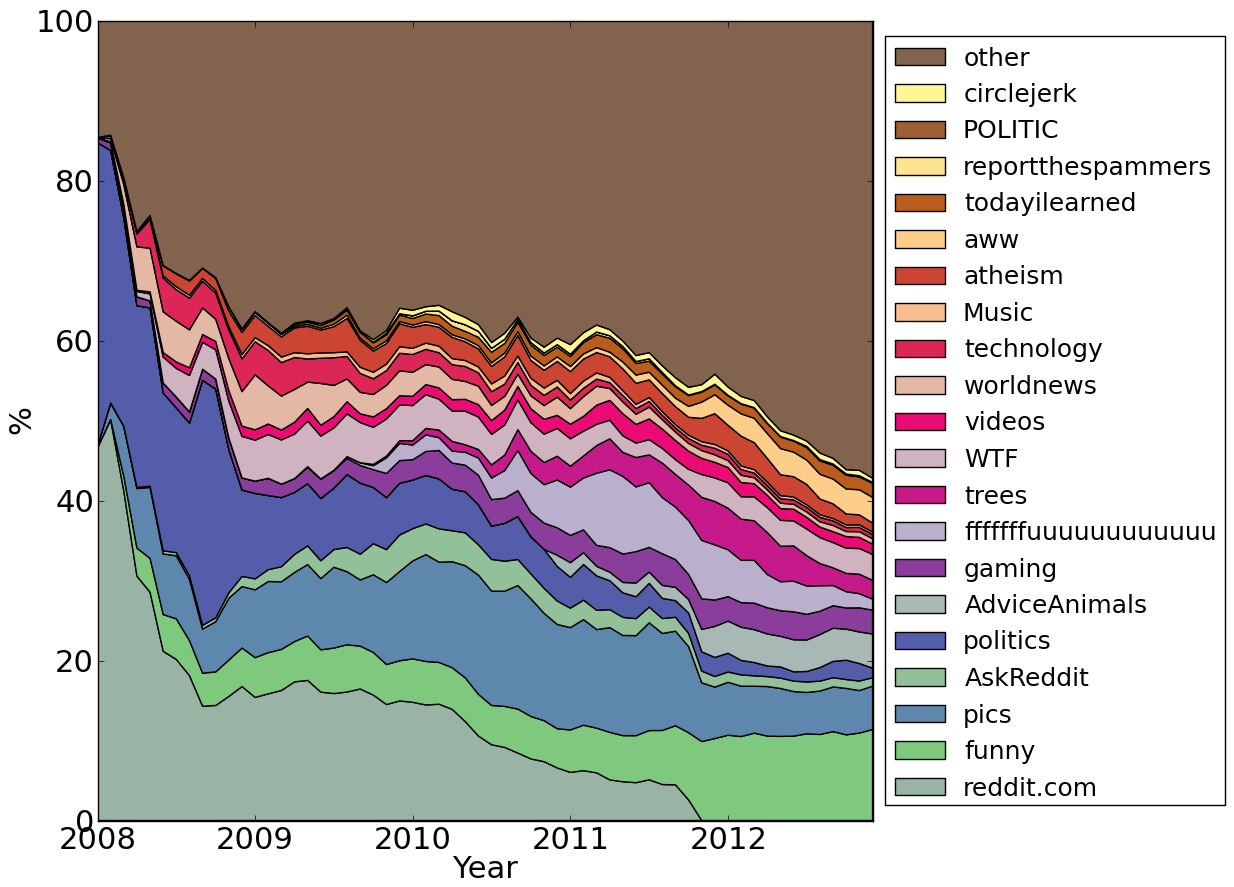}
\caption{Score per subreddit.}
\label{fig:scoresubreddit}
\end{subfigure}
\begin{subfigure}[b]{0.33\textwidth}
\includegraphics[width=\textwidth]{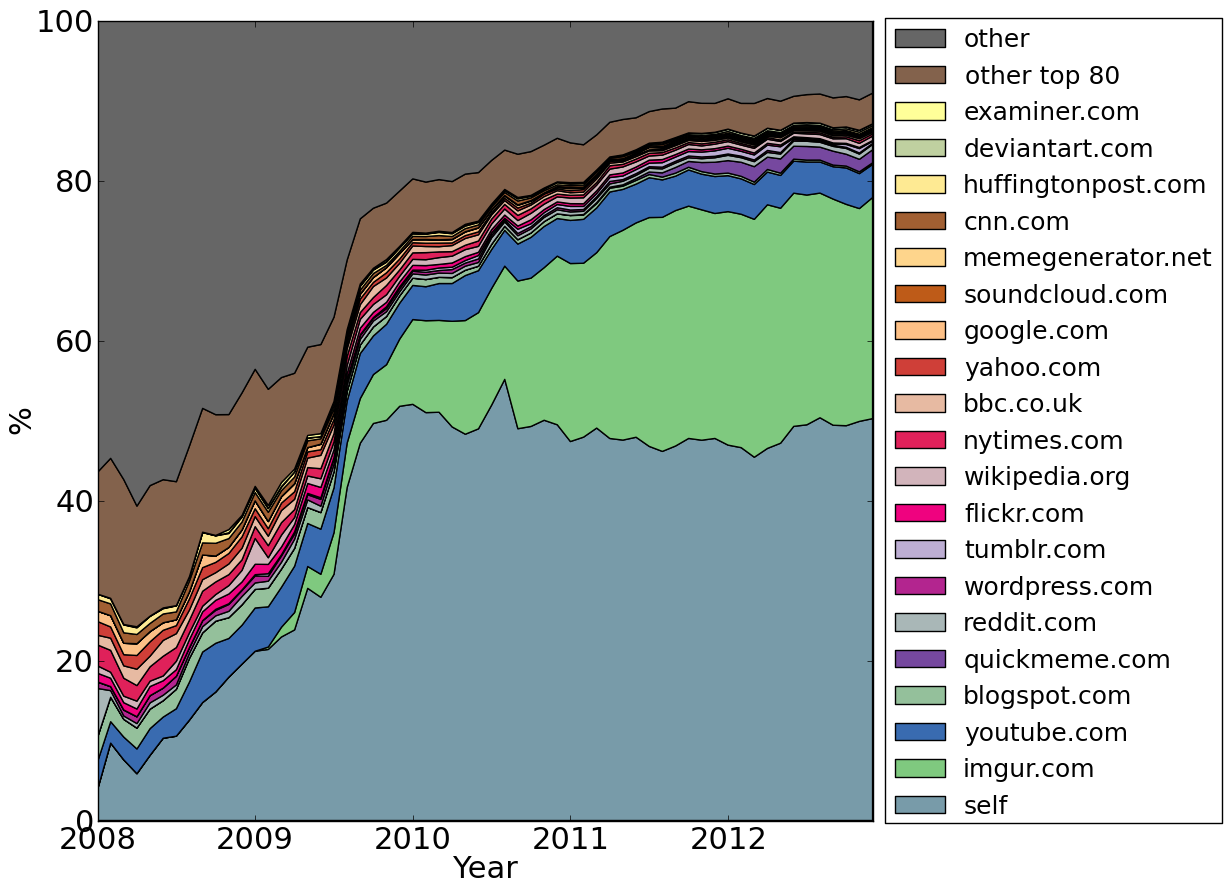}
\caption{Number of comments per domain.}
\label{fig:commentsdomains}
\end{subfigure}
\begin{subfigure}[b]{0.33\textwidth}
\includegraphics[width=\textwidth]{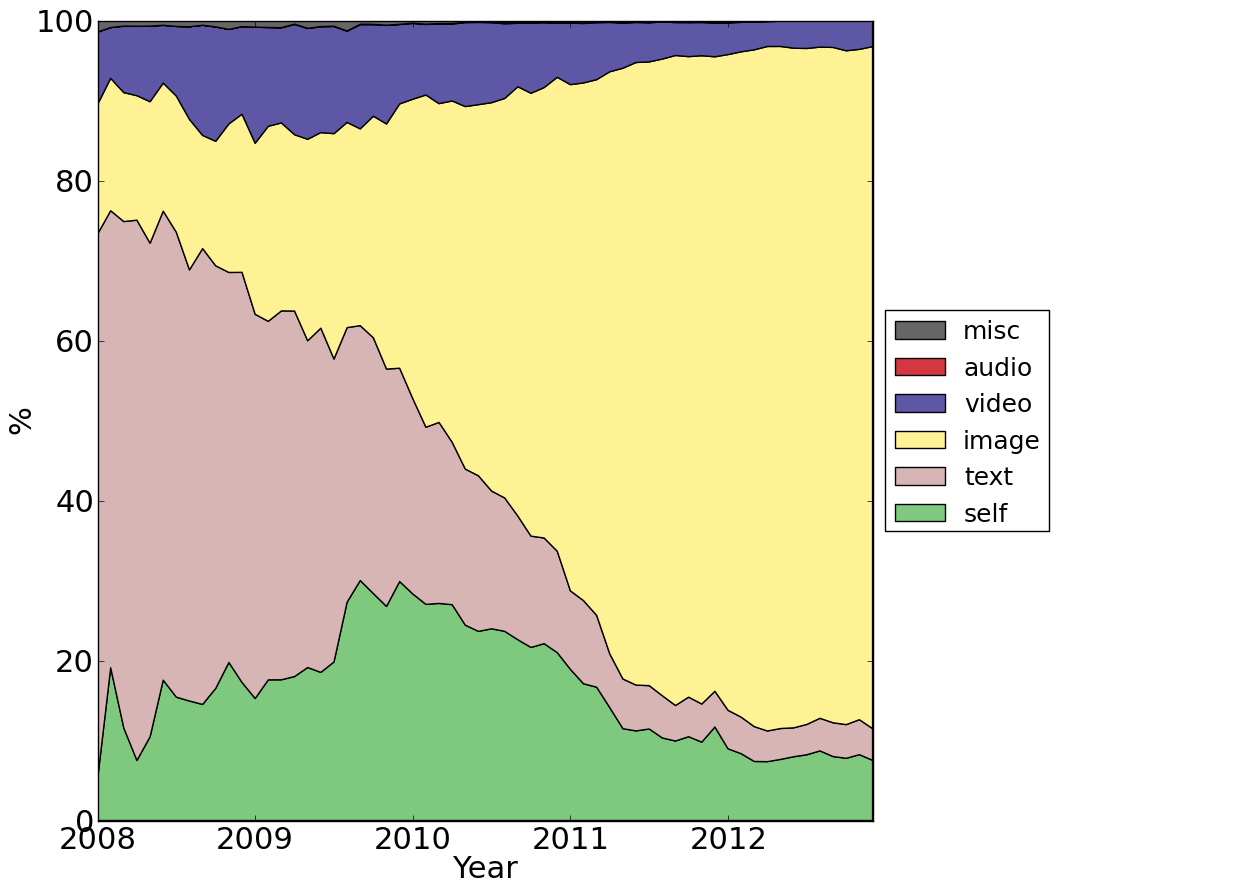}
\caption{Number of votes per category}
\label{subfig:devvot}
\end{subfigure}
\caption{Evolution of score per subreddit, number of comments per domain and number of votes per category, 2008-2012.}
\label{fig:attentionsubdom}
\vspace{-0.2cm}
\end{figure*}

\begin{table}[b!]
\vspace{-0.3cm}
\centering
\caption{User study results 1. Left: Percent of participants agreeing to a description of Reddit (multiple choice). Right: Percentage of participants using features of Reddit (optional questions). Cf. fn. \ref{fn:survey}}
\scriptsize
\begin{tabular}{| l | l |}
\hline 
How would you characterize Reddit? & \% \\ \hline 
Forum / Message board & 88 \\ 
Entertainment site & 71 \\
News site & 56 \\ 
Image/Video or file sharing site & 54  \\ 
Portal & 48 \\ 
Educational site &43  \\ 
Social Network & 33 \\ 
Other & 26 \\ \hline
\end{tabular}
\quad
\begin{tabularx}{0.4\columnwidth}{| X | l |}
\hline 
Users said they... & \% \\ \hline 
...never/seldom submit content to Reddit. (\emph{n=669})& 78 \\ 
...often or very often vote on submissions. (\emph{n=670})& 55 \\
...often or very often comment on submissions. (\emph{n=665})& 32 \\ \hline
\end{tabularx}
\label{tab:survey}
\end{table}

Section \ref{sec:diversity} uncovered that, regarding \emph{submissions}, self-referential content developed a dominating presence over posted links providing a gateway out into the Web. 
Yet, online communities usually display a large discrepancy between the amount of users submitting content and users only consuming content, with the latter number usually being much larger \cite{nonnecke}. Our user survey confirms (at least for our sample) that this also applies to Reddit (see Table~\ref{tab:survey}).


To be able to make statements about these ``lurkers'' as well, we consequently examine whether the \emph{attention} of the Reddit community as a whole follows the emerging focus of offered content and how this content is \emph{perceived}, i.e., appreciated.
We study perception and attention using two basic mechanisms on Reddit: (a) voting and (b) commenting. The \emph{score} -- i.e., upvotes minus downvotes -- provides a proxy for the perception of users towards a submission. The \emph{number of total votes} tells us the overall voting attention users pay to submissions, while the \emph{number of comments} is a proxy for the affinity to discuss content. Our data reveals that Redditors seem to have become less critical over time, indicated by a rising average score per submission. This already hints towards a generally positive perception of Reddit's evolution. In the remainder of this section, we investigate perception and attention of subreddits, linked domains and the type of content over time more in detail.

\subsection{How users view subreddits}
\label{subsec:attsubreddits}

Section~\ref{sec:changesubreddit} revealed a strong, ongoing diversification of subreddits.
Looking at the overall perception and attention of submissions in these distinct subreddits, we can identify interesting patterns. 
At the end of 2012, the top 20 subreddits occupied more than 70\% of all votes casted on submissions on Reddit (with \emph{r/funny} and \emph{r/pics} being the top subreddits in regards to total votes). However, looking at the percentage of the total score of specific subreddits draws a clearer picture as one can see in Figure~\ref{fig:scoresubreddit}. The figure shows that the positive attitude of Redditors -- i.e., the score of submissions -- gets fragmented more and more over different subreddits. The number of comments per subreddit evolved similar, further strengthening our observations. This suggests that the diversification of submitted content over a large number of subreddits over time is coupled with a positive perception of the diversified content. This means Redditors may not necessarily be focused on submissions to top-subreddits only, but rather diversify their interests over a series of distinct sub-communities.

\subsection{How users view external domains}
\label{subsec:attdomains}

Section~\ref{sec:changedomains} demonstrated that a concentration of submissions to a few domains takes place -- mainly self and \emph{Imgur.com}. 
When looking at the number of comments for distinct domains, we can clearly see that users' attention focuses on just a few domains -- again self and Imgur -- over time (depicted in Figure~\ref{fig:commentsdomains}). The growth in comments for Imgur even surpasses what could be expected from the increase in submissions. Still, self submissions are the primary driver of conversations, evident in the large number of comments that can be attributed to self submissions. This is in line with the topics of self submissions we observed in Section \ref{sec:contentchange}, mainly to \emph{r/AskReddit} and \emph{r/circlejerk}.

\subsection{How users view types of content}
\label{subsec:attcontent}


In Section~\ref{sec:contentchange} we found that image and self submissions have become the main type of content submitted to Reddit. Figure~\ref{subfig:devvot} shows how the total number of votes for each submission category (cf. Section~\ref{subsec:dataset}) is distributed over time. Here we can observe that image submissions receive a dominant portion of the total votes (ca. 85\%) at the end of 2012 (up from ca. 16\% at the beginning of 2008). In contrast, votes for self submissions are in slight decline - totaling around 8\% in late 2012. This finding could be reflective of a concern by some Redditors that Reddit has turned itself into an image board. Our results also reveal that self submissions have evolved to attract 50\% of all comments at the end of 2012 (up from ca. 4\% at the beginning of 2008). In other words, an increase of both image as well as self submissions is accompanied by a surge in attention: by a high number of votes for image submissions and a high number of comments for self submissions. This suggests that different types of submissions lend themselves to different types of community reactions, and that these reactions can - sometimes drastically - change over time.

\section{Discussion}
\label{sec:discussion}

While a diversification into thousands of user-generated subreddits has taken place, aided by the demise of \emph{r/Reddit.com}, overall there has been a significant concentration of the domains that submissions are linked to, with Imgur and Reddit itself being the main goals of posts, and to a lesser degree followed by Youtube and Quickmeme. The growth of these domains is aligned with an increase of image and self posts over the years and also indicates an assimilation of other image hosts by Imgur. This development is coupled with a fragmentation of Redditors' overall increasingly positive attitude towards all subreddits as well as a high concentration of attention on just a few domains where submissions link to.
The profound shift of submissions to Reddit itself and Imgur -- which, arguably, was for a long time not more than the image hosting extension of Reddit -- has been running parallel to an overall proliferation of the number of distinct, linked top-level domains with the number of submissions.

The growing dedication to self-hosting lends itself to the idea that a large share of content is both created by the community and addressed at the community, entailing the hypothesis that Reddit has been experiencing an increasing, fundamental shift from ``out-reference'' to more ``self-reference''\footnote{During the review process for this paper, a related discussion has also emerged on Reddit: \url{http://reddit.com/tb/1vaiea}.}.
This assertion is backed by the fact that self submissions have been focused, to an overwhelming degree, on questions by and to the community in the style of a question and answer site (\emph{r/AskReddit}) and to no small part on a subreddit for the sole purpose of being self-ironic about the community (\emph{r/circlejerk}). In these instances, Reddit is clearly occupied with itself. Other popular subreddits discuss external topics in their self posts, but still link out rarely (e.g., \emph{r/politics} and \emph{r/atheism}). 
Hence, there seems to be one part of Reddit based on self posts that is more discussion-, and social-oriented and which became much stronger since 2008. The answers to our survey in Table~\ref{tab:survey} underline this via the perception of Reddit as a forum.

Regarding pictures (hosted mostly on Imgur), we can observe that they are often created by users to convey a personal message\linebreak (\emph{r/AdviceAnimals} and {\emph{r/fu} -- memes from Quickmeme), and argue in other cases that they very frequently contain some personal reference to the user posting them (\emph{r/pics} and \emph{r/funny}). 
While both self and image submissions receive a high amount of attention by Redditors, image submissions gain a much larger portion of the total votes and total score on Reddit, whereas self submissions get commented on more heavily -- however not to the same degree as images surpass self posts in total votes and score.
Both the voting and commenting behavior suggests that the rise in images on Reddit is perceived very positively, while the upsurge in self submissions mostly gets reflected by the amount of comments they receive, generally with a positive attitude. The high amount of votes for images might be explained by the fact that users need to invest much less cognitive load and time to judge them compared to a self (text) post, while the latter is more likely to instigate discussions. 
As submitting Redditors are ever hunting for ``karma'' in the form of upvotes, these preferences certainly encourage to expand the offer on such content, further to be (positively) voted on.
However, we cannot confirm such causality in one way or the other but leave the investigation to future research.  


By and large, coming back to the analogy of the ``frontpage'' of (the best) Internet content that Reddit wants to be, it is not said that it is not a gateway to thousands of diverse parts of the Web (anymore); the number of distinct, linked top-level domains has increased over time, keeping more or less up with the general growth in submissions.
Yet, an overly large portion of the submissions available to users is now taken up by (often personal or community-specific) images and self posts instead of external links, in some cases so particular to Reddit's community (as can be claimed for the cases of ``Rage comics'' and ``Advice Animals''), that only frequent Redditors can fully get the intended meaning.
Compared to the early phase of Reddit, visitors are nowadays much more likely to end up consuming self-referential content instead of finding their way through the proverbial gateway back out into the Web. This is on the one hand because of the probability of said content to make up large parts of ranked submission lists, due to its sheer volume. But it can also be attributed to users' own affinity for such content, which they seek out or are at least prone to vote on and thereby catapulting it up the ``hot'' lists of Reddit, instigating a self-reinforcing cycle. It is also valid to assume that with more exposure to community-centric content, users become more involved in Reddit's ``biotope'', producing such content themselves in turn. The centrality of Reddit for the information diet of its users is underlined by our user study (Table \ref{tab:survey2}), exemplifying that Reddit is (a) the main website for certain topics for many users, (b) is often visited daily and (c) is rather used with no specific information need in mind, leaving users susceptible to the suggestions by the system.

Arguably, the community aspect of Reddit is becoming more important and images and self posts are the prime communication means between its members, an assumption strengthened by the survey results in Table~\ref{tab:survey}, highlighting messaging, entertainment and pictures, while still prominently mentioning news and the portal function.
We can witness the growth and increasing self-definition of a community -- a phenomenon we could only briefly revisit in this paper. 
Without judging this evolution in one way or another, it certainly changes the nature of Reddit as a link-sharing platform, affecting the once straightforward and simple link exchange in ways that merit further study.

\begin{table}[b!]
\vspace{-0.3cm}
\centering
\caption{User study results 2. Left: Percent of participants naming Reddit as their main source of a specific content (multiple choice). Right: Mean answer values for various questions. Cf. fn. \ref{fn:survey}}
\scriptsize
\begin{tabular}{| l | l |}
\hline 
Main website for content? & \% \\ \hline 
Entertainment/Distraction & 90 \\ 
Education, advice, learning & 61 \\
News & 59 \\ 
Social interaction, discussion & 46  \\ 
File sharing & 5 \\ 
Not main site for any content &6  \\ 
Other & 5 \\ \hline
\end{tabular}
\quad
\begin{tabularx}{0.46\columnwidth}{| X | l |}
\hline
Question & Mean\\
\hline 
\# of websites visited daily& 11.80 \\ 
Rank of Reddit among top 10 daily sites& 1.98 \\
Likert 1-7, 1= ``Look\-ing for smth. spec\-ific'', ~~~~~~7= ``Just ex\-ploring'' & 5.27 \\ \hline
\end{tabularx}
\label{tab:survey2}
\vspace{-0.2cm}
\end{table}

\section{Related Work}
\label{sec:related}
Lakkaraju et al.~\cite{lakkaraju} studied how titles, submission times and community choices of image submissions affect the success of the content by investigating resubmitted images on Reddit, showing that good content can speak for itself, although a good title has a positive effect on popularity. Gilbert \cite{gilbert} investigated resubmissions  of content to Reddit and compared their eventual voting score, 
finding that identical links are ignored by the community several times before achieving popularity. Weninger et al. \cite{weninger} focus on comment threads on Reddit, showing that highest scoring comments are mostly submitted at early stages of the discussion.
For the similar platform \emph{digg.com}, studies comparable to the above have been conducted, some  juxtaposing Digg and Reddit in specific aspects (e.g., Lerman \cite{lerman2006social}). In addition, the Reddit community itself has shown an interest in the evolution of the platform. An example can be found in a blog post \cite{Olson2013} that has looked at the evolution of submissions via subbreddits (cf. Figure~\ref{subfig:subredditchange}), suggesting a trend towards diversification. While this blog post presents interesting initial insights into Reddit's development, our work advances them by systematically studying and comparing the evolution of domains, content types, the perception of submissions via scores, comments and votes and other aspects coupled with a user survey (969 respondents) in order to get a more comprehensive understanding of Reddit's evolution.





\section{Conclusions}
\label{sec:conclusions}

To the best of our knowledge, this work represents the most comprehensive longitudinal study of Reddit's evolution to date, studying both (i) how user submissions have evolved over time and (ii) how the community's allocation of attention and its perception of submissions have changed over 5 years based on an analysis of almost 60 million submissions. Our main findings are threefold: \emph{(i) Increasing diversification of topics:} we found that Reddit has evolved from a small community capturing a broad topic area to a platform covering a large number of distinct sub-communities with specialized interests and topics. \emph{(ii) Concentration towards a few domains:} we can observe that over time, submissions and attention (comments) increasingly focused on two domains, i.e., \emph{Imgur.com} and self, i.e. the Reddit community increasingly reinforces its own user-generated image- and textual content over external sources. These results suggest that \emph{(iii) Reddit has transformed from a dedicated gateway to the Web (``The front page of the Internet'') to an increasingly self-referential community.}  Our results are both reflected by how submissions are posted on Reddit as well as how Redditors perceive the ever-changing arrangement of submissions and how they divide their attention among them. From our analysis it remains unclear whether the observed changes in Reddit's community are the result of a conscious effort (e.g., by the operators of the platform or by influential subreddits), or whether the community merely gradually drifted towards a more self-referential mode of operation. We leave answering this question to future work.


Overall, our work shows how an online community with (a) high degrees of freedom for users (e.g., voting, commenting or creating sub-communities) and (b) exceptional growth over several years may dramatically change its nature and focus over time. While we delivered a preliminary analysis of Reddit as an example of a large and growing community, we hope that our work inspires others to expand this line of research to more in-depth studies of Reddit and other comparable community platforms (such as hackernews\footnote{\url{https://news.ycombinator.com/}}).

\smallskip
\textbf{Acknowledgments}
We thank Jason Baumgartner of \emph{Redditanalytics.com} for supplying us with the Reddit log data. This work was partially funded by the FWF Austrian Science Fund Grant I677.

%
\bibliographystyle{abbrv}
\bibliography{sigproc}  

\begin{thebibliography}{1}

\bibitem{gilbert}
E.~Gilbert.
\newblock Widespread underprovision on reddit.
\newblock In {\em Proceedings of the 2013 Conference on Computer Supported
  Cooperative Work}, CSCW '13, pages 803--808, New York, NY, USA, 2013. ACM.

\bibitem{lakkaraju}
H.~Lakkaraju, J.~McAuley, and J.~Leskovec.
\newblock What's in a name? understanding the interplay between titles,
  content, and communities in social media.
\newblock In {\em Seventh International AAAI Conference on Weblogs and Social
  Media}, 2013.

\bibitem{lerman2006social}
K.~Lerman.
\newblock Social networks and social information filtering on digg.
\newblock In {\em Proceedings of International Conference on Weblogs and Social
  Media}, 2007.

\bibitem{nonnecke}
B.~Nonnecke and J.~Preece.
\newblock Lurker demographics: Counting the silent.
\newblock In {\em Proceedings of the SIGCHI Conference on Human Factors in
  Computing Systems}, CHI '00, pages 73--80, New York, NY, USA, 2000. ACM.

\bibitem{Olson2013}
R.~Olson.
\newblock {Retracing the evolution of Reddit through post data,
  http://dx.doi.org/10.6084/m9.figshare.650851}, Mar. 2013.

\bibitem{weninger}
T.~Weninger, X.~A. Zhu, and J.~Han.
\newblock An exploration of discussion threads in social news sites: A case
  study of the reddit community.
\newblock In {\em Proceedings of the 2013 {IEEE}/ACM International Conference
  on Social Networks Analysis and Mining ({ASONAM} 2013)}, Aug. 2013.

\end{thebibliography}
%
%

\end{document}